\pdfoutput=1
\documentclass[12pt,a4paper]{article}
\usepackage{ifthen} % for conditional statements
\newboolean{pdflatex}
\setboolean{pdflatex}{true} % False for eps figures

\newboolean{articletitles}
\setboolean{articletitles}{true} % False removes titles in references

\newboolean{uprightparticles}
\setboolean{uprightparticles}{false} %True for upright particle symbols

\newboolean{inbibliography}
\setboolean{inbibliography}{false} %True once you enter the bibliography

\usepackage{microtype}
\usepackage{lineno}  % for line numbering during review
\usepackage{xspace} % To avoid problems with missing or double spaces after
\usepackage{caption} %these three command get the figure and table captions automatically small

\usepackage{graphicx}  % to include figures (can also use other packages)
\usepackage{color}
\usepackage{colortbl}
\graphicspath{{./figs/}} % Make Latex search fig subdir for figures

\usepackage{amsmath} % Adds a large collection of math symbols
\usepackage{amssymb}
\usepackage{amsfonts}
\usepackage{upgreek} % Adds in support for greek letters in roman typeset

\newcommand*\patchAmsMathEnvironmentForLineno[1]{%
\expandafter\let\csname old#1\expandafter\endcsname\csname #1\endcsname
\expandafter\let\csname oldend#1\expandafter\endcsname\csname
end#1\endcsname
 \renewenvironment{#1}%
   {\linenomath\csname old#1\endcsname}%
   {\csname oldend#1\endcsname\endlinenomath}%
}
\newcommand*\patchBothAmsMathEnvironmentsForLineno[1]{%
  \patchAmsMathEnvironmentForLineno{#1}%
  \patchAmsMathEnvironmentForLineno{#1*}%
}
\AtBeginDocument{%
\patchBothAmsMathEnvironmentsForLineno{equation}%
\patchBothAmsMathEnvironmentsForLineno{align}%
\patchBothAmsMathEnvironmentsForLineno{flalign}%
\patchBothAmsMathEnvironmentsForLineno{alignat}%
\patchBothAmsMathEnvironmentsForLineno{gather}%
\patchBothAmsMathEnvironmentsForLineno{multline}%
\patchBothAmsMathEnvironmentsForLineno{eqnarray}%
}

\usepackage{hyperref}    % Hyperlinks in references
\usepackage[all]{hypcap} % Internal hyperlinks to floats.

\def\lhcb {\mbox{LHCb}\xspace}

\def\MagUp {\mbox{\em Mag\kern -0.05em Up}\xspace}

\ifthenelse{\boolean{uprightparticles}}%
{

 \def\Pmu         {\ensuremath{\upmu}\xspace}

 \def\Ppi         {\ensuremath{\uppi}\xspace}
 
 \def\Prho        {\ensuremath{\uprho}\xspace}

 \def\Ppsi        {\ensuremath{\uppsi}\xspace}

 \def\PDelta      {\ensuremath{\Delta}\xspace}
 \def\PXi      {\ensuremath{\Xi}\xspace}
 \def\PLambda      {\ensuremath{\Lambda}\xspace}
 \def\PSigma      {\ensuremath{\Sigma}\xspace}
 \def\POmega      {\ensuremath{\Omega}\xspace}
 \def\PUpsilon      {\ensuremath{\Upsilon}\xspace}

 \def\PB      {\ensuremath{\mathrm{B}}\xspace}
 
 \def\PD      {\ensuremath{\mathrm{D}}\xspace}

 \def\PJ      {\ensuremath{\mathrm{J}}\xspace}
 \def\PK      {\ensuremath{\mathrm{K}}\xspace}

 \def\Pb      {\ensuremath{\mathrm{b}}\xspace}
 \def\Pc      {\ensuremath{\mathrm{c}}\xspace}
 
 \def\Pe      {\ensuremath{\mathrm{e}}\xspace}

 \def\Pi      {\ensuremath{\mathrm{i}}\xspace}

 \def\Ps      {\ensuremath{\mathrm{s}}\xspace}

}
{

 \def\Pmu         {\ensuremath{\mu}\xspace}

 \def\Ppi         {\ensuremath{\pi}\xspace}
 
 \def\Prho        {\ensuremath{\rho}\xspace}

 \def\Ppsi        {\ensuremath{\psi}\xspace}
 
 \mathchardef\PDelta="7101
 \mathchardef\PXi="7104
 \mathchardef\PLambda="7103
 \mathchardef\PSigma="7106
 \mathchardef\POmega="710A
 \mathchardef\PUpsilon="7107
 
 \def\PB      {\ensuremath{B}\xspace}
 
 \def\PD      {\ensuremath{D}\xspace}

 \def\PJ      {\ensuremath{J}\xspace}
 \def\PK      {\ensuremath{K}\xspace}

 \def\Pb      {\ensuremath{b}\xspace}
 \def\Pc      {\ensuremath{c}\xspace}
 
 \def\Pe      {\ensuremath{e}\xspace}

 \def\Pi      {\ensuremath{i}\xspace}

 \def\Ps      {\ensuremath{s}\xspace}

}

\makeatletter
\ifcase \@ptsize \relax% 10pt
  \newcommand{\miniscule}{\@setfontsize\miniscule{4}{5}}% \tiny: 5/6
\or% 11pt
  \newcommand{\miniscule}{\@setfontsize\miniscule{5}{6}}% \tiny: 6/7
\or% 12pt
  \newcommand{\miniscule}{\@setfontsize\miniscule{5}{6}}% \tiny: 6/7
\fi
\makeatother

\DeclareRobustCommand{\optbar}[1]{\shortstack{{\miniscule (\rule[.5ex]{1.25em}{.18mm})}
  \\ [-.7ex] $#1$}}

\def\en         {{\ensuremath{\Pe^-}}\xspace}   % electron negative (\em is taken)
\def\ep         {{\ensuremath{\Pe^+}}\xspace}

\def\mup        {{\ensuremath{\Pmu^+}}\xspace}
\def\mun        {{\ensuremath{\Pmu^-}}\xspace} % muon negative (\mum is taken)

\def\squark    {{\ensuremath{\Ps}}\xspace}

\def\cquark    {{\ensuremath{\Pc}}\xspace}

\def\bquark    {{\ensuremath{\Pb}}\xspace}

\def\pion   {{\ensuremath{\Ppi}}\xspace}
\def\piz    {{\ensuremath{\pion^0}}\xspace}

\def\pip    {{\ensuremath{\pion^+}}\xspace}
\def\pim    {{\ensuremath{\pion^-}}\xspace}

\def\rhomeson {{\ensuremath{\Prho}}\xspace}

\def\rhop     {{\ensuremath{\rhomeson^+}}\xspace}

\def\kaon    {{\ensuremath{\PK}}\xspace}
  \def\Kbar    {{\kern 0.2em\overline{\kern -0.2em \PK}{}}\xspace}

\def\KorKbar    {\kern 0.18em\optbar{\kern -0.18em K}{}\xspace}

\def\Kp      {{\ensuremath{\kaon^+}}\xspace}

\def\KS      {{\ensuremath{\kaon^0_{\rm\scriptscriptstyle S}}}\xspace}

\def\Kstarz  {{\ensuremath{\kaon^{*0}}}\xspace}

\def\Kstarp  {{\ensuremath{\kaon^{*+}}}\xspace}

  \def\Dbar    {{\kern 0.2em\overline{\kern -0.2em \PD}{}}\xspace}

\def\DorDbar    {\kern 0.18em\optbar{\kern -0.18em D}{}\xspace}

\def\B       {{\ensuremath{\PB}}\xspace}
\def\Bbar    {{\ensuremath{\kern 0.18em\overline{\kern -0.18em \PB}{}}}\xspace}

\def\BorBbar    {\kern 0.18em\optbar{\kern -0.18em B}{}\xspace}
\def\Bz      {{\ensuremath{\B^0}}\xspace}

\def\Bu      {{\ensuremath{\B^+}}\xspace}

\def\Bp      {{\ensuremath{\Bu}}\xspace}

\def\Bd      {{\ensuremath{\B^0}}\xspace}
\def\Bs      {{\ensuremath{\B^0_\squark}}\xspace}

\def\jpsi     {{\ensuremath{{\PJ\mskip -3mu/\mskip -2mu\Ppsi\mskip 2mu}}}\xspace}

  \def\Y#1S{\ensuremath{\PUpsilon{(#1S)}}\xspace}% no space before {...}!

\def\Lbar        {{\ensuremath{\kern 0.1em\overline{\kern -0.1em\PLambda}}}\xspace}
\def\LorLbar    {\kern 0.18em\optbar{\kern -0.18em \PLambda}{}\xspace}

\def\BF         {{\ensuremath{\cal B}}\xspace}

\newcommand{\decay}[2]{\ensuremath{#1\!\to #2}\xspace}         % {\Pa}{\Pb \Pc}

\def\to                 {\ensuremath{\rightarrow}\xspace}

\def\CP                {{\ensuremath{C\!P}}\xspace}

\def\AT#1     {\ensuremath{A_{\mathrm{T}}^{#1}}\xspace}           % 2

\def\C#1      {\ensuremath{\mathcal{C}_{#1}}\xspace}                       % 9
\def\Cp#1     {\ensuremath{\mathcal{C}_{#1}^{'}}\xspace}                    % 7
\def\Ceff#1   {\ensuremath{\mathcal{C}_{#1}^{\mathrm{(eff)}}}\xspace}        % 9
\def\Cpeff#1  {\ensuremath{\mathcal{C}_{#1}^{'\mathrm{(eff)}}}\xspace}       % 7
\def\Ope#1    {\ensuremath{\mathcal{O}_{#1}}\xspace}                       % 2
\def\Opep#1   {\ensuremath{\mathcal{O}_{#1}^{'}}\xspace}                    % 7

\newcommand{\tev}{\ifthenelse{\boolean{inbibliography}}{\ensuremath{~T\kern -0.05em eV}\xspace}{\ensuremath{\mathrm{\,Te\kern -0.1em V}}}\xspace}
\newcommand{\gev}{\ensuremath{\mathrm{\,Ge\kern -0.1em V}}\xspace}
\newcommand{\mev}{\ensuremath{\mathrm{\,Me\kern -0.1em V}}\xspace}
\newcommand{\kev}{\ensuremath{\mathrm{\,ke\kern -0.1em V}}\xspace}
\newcommand{\ev}{\ensuremath{\mathrm{\,e\kern -0.1em V}}\xspace}
\newcommand{\gevc}{\ensuremath{{\mathrm{\,Ge\kern -0.1em V\!/}c}}\xspace}
\newcommand{\mevc}{\ensuremath{{\mathrm{\,Me\kern -0.1em V\!/}c}}\xspace}
\newcommand{\gevcc}{\ensuremath{{\mathrm{\,Ge\kern -0.1em V\!/}c^2}}\xspace}
\newcommand{\gevgevcccc}{\ensuremath{{\mathrm{\,Ge\kern -0.1em V^2\!/}c^4}}\xspace}
\newcommand{\mevcc}{\ensuremath{{\mathrm{\,Me\kern -0.1em V\!/}c^2}}\xspace}

\def\mum  {\ensuremath{{\,\upmu\rm m}}\xspace}

\def\invfb   {\ensuremath{\mbox{\,fb}^{-1}}\xspace}

\def\gsim{{~\raise.15em\hbox{$>$}\kern-.85em
          \lower.35em\hbox{$\sim$}~}\xspace}
\def\lsim{{~\raise.15em\hbox{$<$}\kern-.85em
          \lower.35em\hbox{$\sim$}~}\xspace}

\def\sPlot{\mbox{\em sPlot}\xspace}
\def\sqs   {\ensuremath{\protect\sqrt{s}}\xspace}

\def\ptot       {\mbox{$p$}\xspace}
\def\pt         {\mbox{$p_{\rm T}$}\xspace}

\def\evtgen     {\mbox{\textsc{EvtGen}}\xspace}

\def\geant      {\mbox{\textsc{Geant4}}\xspace}

\def\photos     {\mbox{\textsc{Photos}}\xspace}

\def\pythia     {\mbox{\textsc{Pythia}}\xspace}

\def\tell1  {TELL1\xspace}
\def\ukl1   {UKL1\xspace}

\usepackage{cite} % Allows for ranges in citations
\usepackage{mciteplus}
\usepackage{longtable} % only for template; not usually to be used in PAPERs
\usepackage{multirow}

\begin{document}

\def\BdJpsiPiz{\decay{\Bz}{ \jpsi \piz}}
\def\BsJpsiGamma{\decay{\Bs}{ \jpsi \gamma}}
\def\BdJpsiGamma{\decay{\Bz}{ \jpsi \gamma}}
\newcommand{\Bds}{\ensuremath{\B^{0}_{(s)}}\xspace}
\def\BdsJpsiGamma{ \decay{\Bds} { \jpsi\gamma}}
\def\BdKstarGamma{\decay{\Bd}{ \Kstarz \gamma}}
\newcommand{\obslimitn}{\ensuremath{7.3 \times 10^{-6}}\xspace}
\newcommand{\obslimitnf}{\ensuremath{8.7 \times 10^{-6}}\xspace}
\newcommand{\explimitn}{\ensuremath{5.6 \times 10^{-6}}\xspace}
\newcommand{\explimitnf}{\ensuremath{6.8 \times 10^{-6}}\xspace}
\newcommand{\obslimit}{\ensuremath{7.3\;(8.7)\times 10^{-6}}\xspace}
\newcommand{\cls}{\ensuremath{\mathrm{CL}_{\mathrm{S}}}\xspace}

%%%%%%%%%%%%%%%%%%%%%%%%%
%%%%% Title     %%%%%%%%%
%%%%%%%%%%%%%%%%%%%%%%%%%
\renewcommand{\thefootnote}{\fnsymbol{footnote}}
\setcounter{footnote}{1}

% %%%%%%% CHOOSE TITLE PAGE--------
%%%%%%%%%%%%%%%%%%%%%%%%%
%%%%%  TITLE PAGE  %%%%%%
%%%%%%%%%%%%%%%%%%%%%%%%%
\begin{titlepage}
\pagenumbering{roman}

% Header ---------------------------------------------------
\vspace*{-1.5cm}
\centerline{\large EUROPEAN ORGANIZATION FOR NUCLEAR RESEARCH (CERN)}
\vspace*{1.5cm}
\noindent
\begin{tabular*}{\linewidth}{lc@{\extracolsep{\fill}}r@{\extracolsep{0pt}}}
\ifthenelse{\boolean{pdflatex}}% Logo format choice
{\vspace*{-2.7cm}\mbox{\!\!\!\includegraphics[width=.14\textwidth]{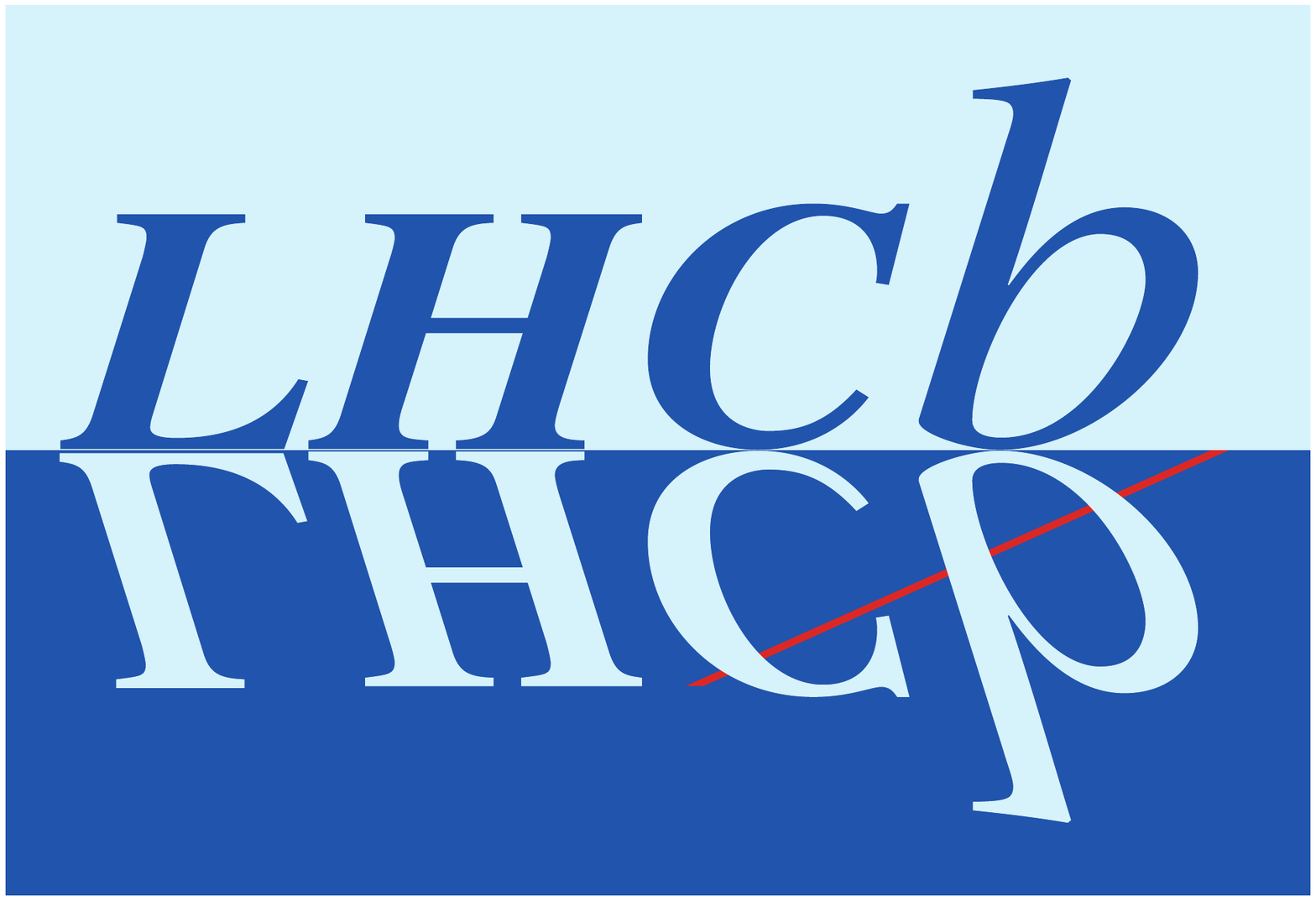}} & &}%
{\vspace*{-1.2cm}\mbox{\!\!\!\includegraphics[width=.12\textwidth]{lhcb-logo.eps}} & &}%
\\
 & & CERN-PH-EP-2015-276 \\  % ID
 & & LHCb-PAPER-2015-044 \\  % ID
 & & 16 October 2015 \\ %\today \\ % Date - Can also hardwire e.g.: 23 March 2010
 & & \\
% not in paper \hline
\end{tabular*}

\vspace*{4.0cm}

% Title --------------------------------------------------
{\bf\boldmath\huge
\begin{center}
  Search for the rare decays $B^{0}\to\jpsi\gamma$ and $B^0_{s}\to\jpsi\gamma$
\end{center}
}

\vspace*{2.0cm}

% Authors -------------------------------------------------
\begin{center}
%In the footnote, replace 'paper' by 'letter' in case of submission to PRL or PLB
LHCb Collaboration\footnote{Full author list given at the end of the paper.}
\end{center}

\vspace{\fill}

% Abstract -----------------------------------------------
\begin{abstract}
  \noindent
  A search for the rare decay of a \Bz or \Bs meson into the final state $\jpsi\gamma$ is
  performed, using data collected by the LHCb experiment in $pp$ collisions at $\sqs=7$ and $8$~TeV,
  corresponding to an integrated luminosity of 3\invfb. The observed number of signal
 candidates is consistent with a background-only hypothesis.
 Branching fraction values larger than
 \mbox{$1.5\times 10^{-6}$} for the $\Bz\to\jpsi\gamma$ decay mode
 are excluded at 90\%~confidence level. For the $\Bs\to\jpsi\gamma$ decay mode,
 branching fraction values larger
 than \mbox{$7.3\times 10^{-6}$} are excluded
 at 90\%~confidence level;
 this is the first branching fraction limit for this decay.
\end{abstract}

\vspace*{2.0cm}

\begin{center}
  Published in Phys.~Rev.~D
\end{center}

\vspace{\fill}

{\footnotesize
\centerline{\copyright~CERN on behalf of the \lhcb collaboration, licence \href{http://creativecommons.org/licenses/by/4.0/}{CC-BY-4.0}.}}
\vspace*{2mm}

\end{titlepage}

%%%%%%%%%%%%%%%%%%%%%%%%%%%%%%%%
%%%%%  EOD OF TITLE PAGE  %%%%%%
%%%%%%%%%%%%%%%%%%%%%%%%%%%%%%%%

%  empty page follows the title page ----
\newpage
\setcounter{page}{2}
\mbox{~}
\cleardoublepage
% %%%%%%%%%%%%% ---------

\renewcommand{\thefootnote}{\arabic{footnote}}
\setcounter{footnote}{0}

%%%%%%%%%%%%%%%%%%%%%%%%%%%%%%%%
%%%%%  Table of Content   %%%%%%
%%%%%%%%%%%%%%%%%%%%%%%%%%%%%%%%
%%%% Uncomment next 2 lines if desired
%\tableofcontents
%\cleardoublepage

%%%%%%%%%%%%%%%%%%%%%%%%%
%%%%% Main text %%%%%%%%%
%%%%%%%%%%%%%%%%%%%%%%%%%

\pagestyle{plain} % restore page numbers for the main text
\setcounter{page}{1}
\pagenumbering{arabic}

%% Uncomment during review phase.
%% Comment before a final submission.
%\linenumbers

% You can include short sections directly in the main tex file.
% However, for larger papers it is desirable to split the text into
% several semiautonomous files, which can be revised independently.
% This is especially useful when developing a document in
% collaboration with several people, since then different parts can be
% edited independently.  This type of file organization is shown here.
%

%\input{introduction}
\section{Introduction}
\label{sec:Introduction}

Decays of \B mesons provide an interesting laboratory to study quantum chromodynamics (QCD).
A typical approach for predicting the branching fractions of such decays is to factorize the decay into a short-distance
contribution which can be computed with perturbative QCD and a long-distance contribution for which
nonperturbative QCD is required. The extent to which this factorization assumption is valid leads to large
theoretical uncertainties. Experimental measurements are therefore crucial to test the different calculations
 of the QCD interactions within these decays, so helping to identify the most appropriate theoretical
 approaches for predicting observables.

\begin{figure}[b]
  \centering
  \includegraphics[width=0.55\textwidth]{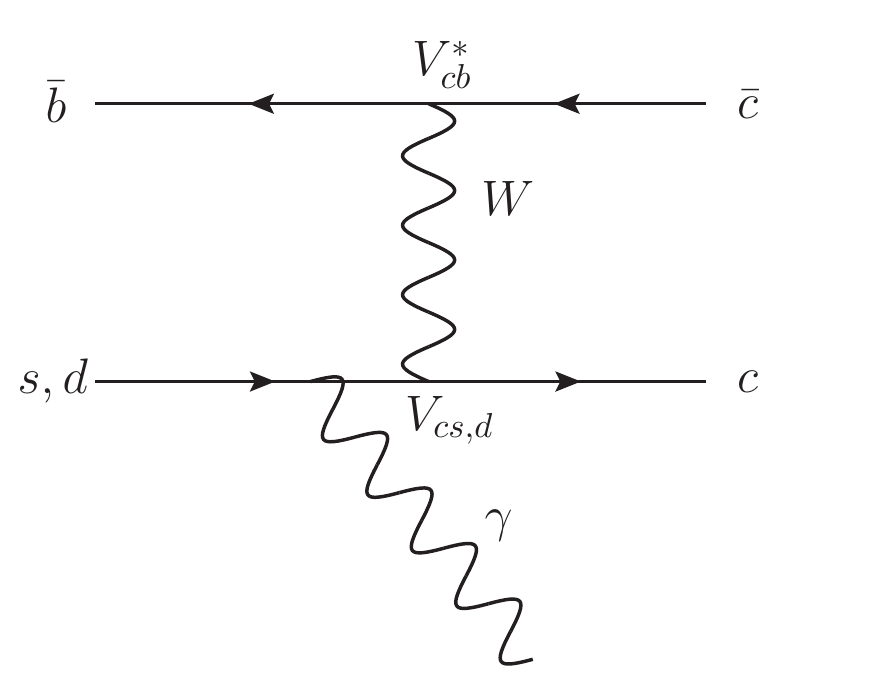}
  \caption{Feynman diagram of the leading contribution to the decay $\B_{(s)}\to\jpsi\gamma$ in the Standard Model. Radiation of the photon from the other quarks is suppressed by a factor of $\Lambda_{\rm{QCD}}/m_{b}$~\cite{Lu:2003ix}.}
  \label{fig:feyn}
\end{figure}

In the SM,
the decays \BdsJpsiGamma proceed through a $W$ boson exchange diagram as shown in Fig.~\ref{fig:feyn},
where one quark radiates a photon (the inclusion of charge conjugate processes is implied
throughout).
 Theoretical predictions of the branching fractions
 of these decays vary significantly depending on the chosen approach for the
 treatment of QCD interactions in the decay dynamics. For example, in Ref.~\cite{Lu:2003ix}
  the branching fraction, evaluated in the framework of QCD factorization~\cite{Collins:1989gx}, is expected to be $\sim 2\times10^{-7}$
  , whereas the calculation in Ref.~\cite{Li:2006xe}, using perturbative
  QCD, predicts a branching fraction of $5\times10^{-6}$. The process is also sensitive to
  physics beyond the SM, for example right-handed currents~\cite{Lu:2003ix}. The decay \BdJpsiGamma has been previously searched for
by the $BABAR$ collaboration, and a limit on the branching fraction of $1.6\times10^{-6}$ was set at 90\% confidence level (C.L.)~\cite{Aubert:2004xd}.

This paper describes a search for the decays \BsJpsiGamma and \BdJpsiGamma,
performed with proton-proton ($pp$) collision data collected by the LHCb experiment
 corresponding to an integrated luminosity of 1.0~(2.0)~\invfb recorded at center-of-mass energies of $\sqs=7$~(8)~TeV.

Event selections are described in Sec.~\ref{sec:selection}.
The signal yield is normalized to a set of $\B\to\jpsi\gamma X$ decays, described in Sec.~\ref{sec:massfit}.
The relative efficiency between
 signal and normalization decay modes is calculated using simulated events. This efficiency
 is cross-checked using the decay $\Bd\to\Kstarz\gamma$. Finally, in Sec.~\ref{sec:results}, the upper limits on the
 branching fractions are calculated using the \cls method~\cite{Read:2002,Junk:1999}.

\section{Detector and simulation}
\label{sec:Detector}

The \lhcb detector~\cite{Alves:2008zz,LHCb-DP-2014-002} is a single-arm forward
spectrometer covering the \mbox{pseudorapidity} range $2<\eta <5$,
designed for the study of particles containing \bquark or \cquark
quarks. The detector includes a high-precision tracking system
consisting of a silicon-strip vertex detector surrounding the $pp$
interaction region, a large-area silicon-strip detector located
upstream of a dipole magnet with a bending power of about
$4{\rm\,Tm}$, and three stations of silicon-strip detectors and straw
drift tubes placed downstream of the magnet.
The combined tracking system provides a measurement of momentum, \ptot, of charged particles with
a relative uncertainty that varies from 0.5\% at low momentum to 1.0\% at 200\gevc.
The minimum distance of a track to a primary vertex, the impact parameter (IP), is measured with a resolution of $(15+29/\pt)\mum$,
where \pt is the component of the momentum transverse to the beam, in\,\gevc.
Different types of charged hadrons are distinguished using information
from two ring-imaging Cherenkov detectors.
Photons, electrons and hadrons are identified by a calorimeter system consisting of
scintillating-pad and preshower detectors, an electromagnetic
calorimeter (ECAL) and a hadronic calorimeter.
Muons are identified by a
system composed of alternating layers of iron and multiwire
proportional chambers.
The online event selection is performed by a trigger system~\cite{LHCb-DP-2012-004},
which consists of a hardware stage, based on information from the calorimeter and muon
systems, followed by a software stage, which applies a full event
reconstruction.

In the simulation, $pp$ collisions are generated using
\pythia~6 and \pythia~8~\cite{Sjostrand:2006za}
 with a specific \lhcb
configuration~\cite{LHCb-PROC-2010-056}.  Decays of hadronic particles
are described by \evtgen~\cite{Lange:2001uf}, in which final-state
radiation is generated using \photos~\cite{Golonka:2005pn}. The
interaction of the generated particles with the detector, and its response,
are implemented using the \geant
toolkit~\cite{Allison:2006ve, *Agostinelli:2002hh} as described in
Ref.~\cite{LHCb-PROC-2011-006}.

\section{Selection requirements}
\label{sec:selection}

Candidate events are first required to pass the hardware trigger
  which requires at least one muon with $\pt>1.48$~(1.76)~\gevc
  in the 7~(8)~\tev data.
  In the subsequent software trigger, at least
  one of the final-state particles is required to have
  $\pt>0.8\gevc$ and IP larger than $100\mum$ with respect to any
  of the primary $pp$ interaction vertices~(PVs) in the
  event. Finally, the tracks of two final-state
  particles are required to form a vertex that is significantly
  displaced from the PVs.

In the offline selection of signal candidates, \jpsi decays are reconstructed from oppositely charged muon pairs where both muons have $\pt >550$\mevc,
good track fit qualities and an IP with respect to any PV significantly different from zero. The muon pair is required to form a good quality decay vertex.
In order to suppress background from decays
 such as $\B\to\jpsi\piz$, where
both photons from the \piz decay are reconstructed as a single cluster in the ECAL, only photons which convert into \ep\en pairs are used in the analysis.
This reduces the signal efficiency by about a factor of 30 with respect
to photons which do not convert, but improves the signal resolution in the reconstructed invariant $\jpsi\gamma$ mass by a factor of 5.
 Furthermore, the direction of the photon momentum vector can be checked for
 consistency with the \B decay vertex and used to reject combinatorial background.
Photons are required to convert in the material before the second tracking
system, which corresponds to about 0.25 radiation lengths~\cite{Alves:2008zz}.
They are reconstructed following a similar strategy to that
described in Ref.~\cite{LHCb-PAPER-2013-028},
by combining electron and positron track pairs, which can be associated with
electromagnetic clusters in the ECAL and are significantly displaced from
the reconstructed $B$ decay vertex.
The energy loss of electrons by emission of bremsstrahlung photons is recovered by
adding the energies of reconstructed photons associated
with the track. The photon candidates are required to have a reconstructed invariant mass less than 100\mevcc and
$\pt>1\gevc$. Candidates are separated into two categories based on where the photon converts in the detector.
Conversions which occur early enough for the converted electrons to be reconstructed in the vertex detector,
are referred to as \emph{long} because the tracks pass through the full tracking system,
while those which convert late enough such that track segments of the
electrons cannot be formed in the vertex detector are referred to as \emph{downstream}~\cite{LHCb-DP-2013-002}. The \jpsi and $\gamma$ candidates are combined to form a \B candidate,
which is required to have an invariant mass in the range $4500<m(\jpsi \gamma)<7000$\mevcc and
$\pt>5\gevc$.
The momentum vector of the \B candidate is required to be aligned with the vector between the associated PV and the decay vertex, in order to
suppress combinational background.

A boosted decision tree~(BDT)~\cite{Breiman,AdaBoost} is trained to reject combinatorial background, where the \jpsi and photon candidates originate from different decays.
The signal is represented in the BDT training with simulated $\Bs\to\jpsi\gamma$ decays, while selected data events
in the high mass sideband, $5500<m(\jpsi \gamma)<6500$~\mevcc, are used to represent the background. The input variables
used in the training are mostly kinematic and geometric variables, as well as isolation criteria used to reject
background containing additional tracks in close proximity to the \jpsi vertex. Separate trainings are performed for events in which the
photon conversion is long or downstream. The $k$-fold cross-validation method~\cite{Kohavi95astudy}, with $k=5$, is used to increase the training
statistics while avoiding overtraining. The requirement on the BDT response is optimized by maximizing the metric
$N_{S}/\sqrt{N_{S}+N_{B}}$, where $N_{S}$ is the estimated number of signal events after selection assuming a
branching fraction $\BF(\Bs\to\jpsi\gamma)=5\times 10^{-6}$, and $N_{B}$ is the estimated number of background
events in the signal region, $5250 < m(\jpsi\gamma) < 5400$, extrapolated from an exponential fit to the data in the high mass sideband.
This requirement is 60\% efficient for simulated signal candidates and rejects 98\% of the combinatorial background.

The decay \BdKstarGamma, where $\Kstarz\to\Kp\pim$, is used to validate the selection and to assess systematic uncertainties arising from differences between simulation and data.
The same BDT used for the signal selection,
with the \jpsi and muon properties replaced by those of the $\Kstarz$ and its decay products,
is applied to the \BdKstarGamma candidates.

\section{Branching fraction}
\label{sec:massfit}

The branching fraction is determined by performing a fit to the $\jpsi\gamma$ invariant mass distribution in the range
$4500<m(\jpsi \gamma)<7000$\mevcc. In the fit, the signal yield is
normalized to the following set of decay modes: $\B^{0}_{(s)}\to\jpsi \eta (\eta\to\gamma\gamma)$, \Bz\to\jpsi\piz,
$\Bz\to\jpsi \KS (\KS\to\piz\piz)$ and
 $\Bp\to\jpsi \rhop (\rhop\to\piz\pip)$, where only the \jpsi meson and one photon are reconstructed.
 These decay modes are chosen because they have relatively well measured branching fractions
 and are expected to contribute in the selected mass range. The normalization procedure is performed by expressing the branching fraction, $\mathcal{B}$, as

\begin{equation}
  \mathcal{B}(B^{0}_{(s)}\to\jpsi\gamma) = \frac{N_{\rm{sig}}}{f_{\rm{sig}}\epsilon_{\rm{sig}}} \,  \frac{\displaystyle\sum_{i} f^{i} \mathcal{B}_{\rm{norm}}^{i} \epsilon^{i}_{\rm{norm}}}{\displaystyle\sum_{i} N^{i}_{\rm{norm}}},
\end{equation}

\noindent where $i$ represents a normalization decay mode, $N_{\rm{sig}}$ and $N^{i}_{\rm{norm}}$ are the observed number of signal and normalization candidates, $f$ is the relevant production fraction and $\epsilon$ is the efficiency as determined from the simulation. Systematic uncertainties associated with these quantities are included in the fit as nuisance parameters.

In general, the normalization modes have a lower offline selection efficiency than
 the signal because the photon has a lower momentum and therefore the electron tracks  are more likely to be bent outside
 the detector acceptance by the magnetic field. For example, the selection efficiency for signal is around 60\% whereas
that of $\Bz\to\jpsi\piz$ is only 30\%.

The dimuon mass is constrained to the known value of the \jpsi meson~\cite{PDG2014},
which improves the $m(\jpsi\gamma)$ resolution by $\sim$30\%.
The \BsJpsiGamma signal shape is obtained by fitting a Gaussian function
with a power-law tail to simulation. The $m(\jpsi\gamma)$ resolution is approximately
90~(70)~\mevcc for long (downstream) decays. The search for
the \BdJpsiGamma signal is performed separately to the \BsJpsiGamma decay, where the same signal shape is
used with the peak position adjusted by the difference in masses of the \Bd and \Bs mesons given in Ref.~\cite{PDG2014}.
The \BdJpsiGamma branching fraction is assumed to be zero when fitting for the \BsJpsiGamma signal, and vice versa.

The normalization modes form a broad shoulder below the signal peak and their shapes are modeled
using dedicated simulation samples. The total normalization yield is allowed to float in the fit,
with the contribution from each individual normalization decay mode constrained, taking into account the
relative efficiencies and branching fractions between them. For the \Bs modes,
the ratio of fragmentation fractions, $f_{s}/f_{d}$ is used to calculate the relative
expected yields of \Bz and \Bs meson decays. The fragmentation fractions
for \Bp and \Bz decays are assumed to be the same.

Other $\B\to\jpsi \gamma X$ decays
which are missing either a heavy particle or several particles are modeled
by an exponential function with the shape obtained from simulated $\Bp\to\jpsi \Kstarp$ events.
The choice of parameterization for these backgrounds is checked
using simulation samples and no bias is observed for the signal yield.
Finally, combinatorial background is modeled by an exponential function,
the slope of which is allowed to float in the fit.

\begin{figure}[tb]
  \centering
  \includegraphics[width=0.8\textwidth]{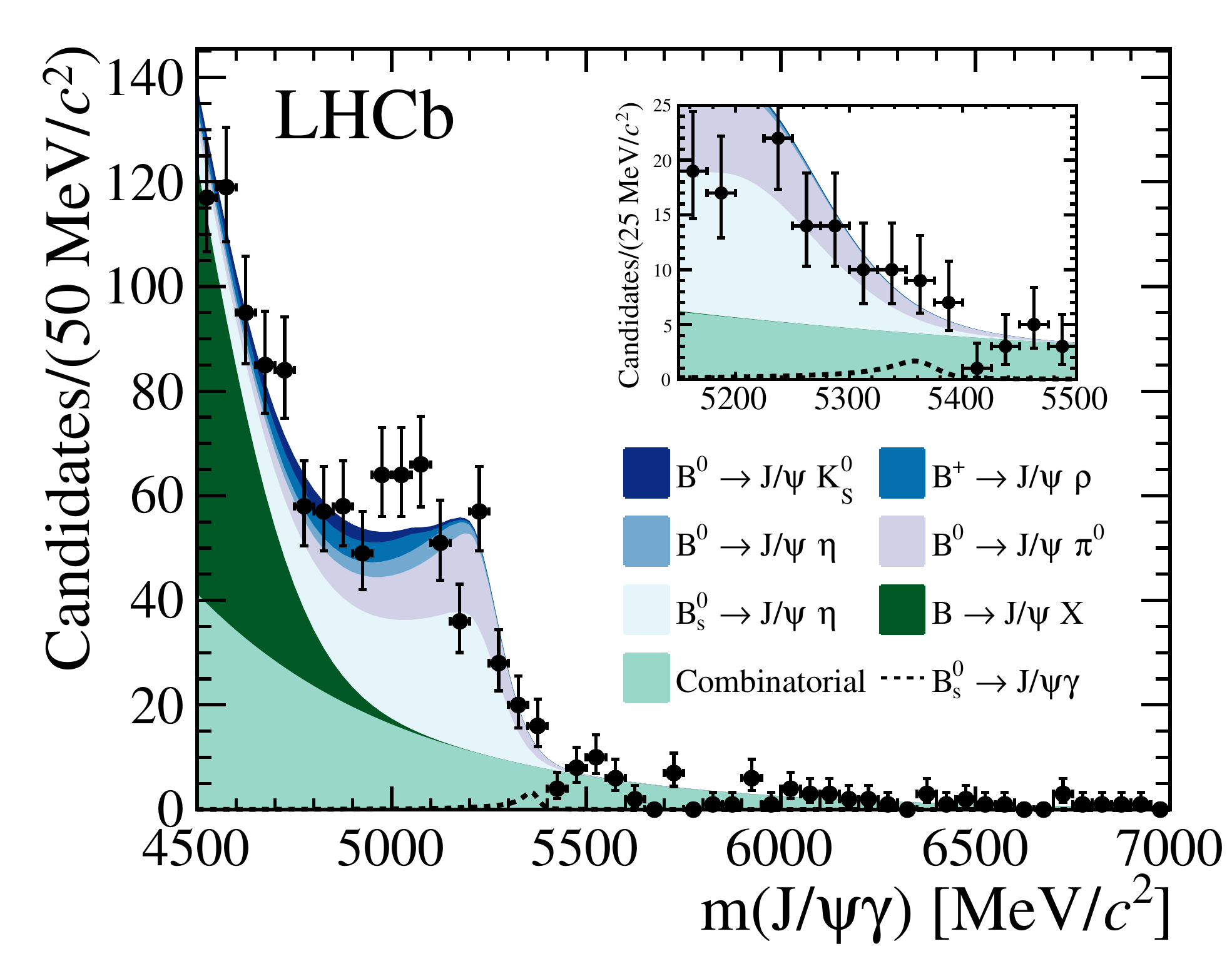}
  \caption{Mass distribution of signal candidates with the fit result overlaid, allowing for a \BsJpsiGamma signal.}
  \label{fig:massfit}
\end{figure}

The result of the fit, to the combined long and downstream samples, allowing for a \BsJpsiGamma contribution, is shown in Fig.~\ref{fig:massfit}, where no
significant signal is observed. The result is similar for the \BdJpsiGamma case.

\section{Systematic uncertainties}
\label{sec:Systematics}

\begin{table}[t]

\newcommand\Tstrut{\rule{0pt}{2.4ex}}
\newcommand\TTstrut{\rule{0pt}{3.2ex}}
\newcommand\Bstrut{\rule[-1.2ex]{0pt}{0pt}}
\newcommand\BBstrut{\rule[-1.8ex]{0pt}{0pt}}
\centering
\caption{Summary of systematic uncertainties. Each source is listed with the corresponding uncertainty on the branching fraction for the signal.}
      \label{tab:system}
     % \resizebox{0.47\textwidth}{!} {
    \begin{tabular}{l c}
  %  \hline
      Source & \hspace{0.2cm}Uncertainty (\%)  \\
      \hline
      Normalisation branching fractions & $\pm17$ \\
      $f_{s}/f_{d}$ & $\pm3$ \\
      \Bs~\CP content & $\pm6$ \\
      Signal shape & $\pm4$ \\
      Simulation mis-modelling & $\pm2$ \\
  %    \hline

    \end{tabular}% }
\end{table}

Many systematic uncertainties cancel to a large extent as both signal and normalization modes contain the same reconstructed final-state particles.
In particular, systematic uncertainties related to the ratio of efficiencies for the trigger and particle identification
requirements are negligible. However, differences can arise as the photon is typically softer for the normalization modes than for the signal.
These effects are accounted for using the $\Bd\to\Kstarz\gamma$ decay as a control channel. All systematics uncertainties are included in the
final likelihood fit as nuisance parameters. Their impact on the branching fraction measurement is summarized in Table.~\ref{tab:system}.

The largest systematic uncertainty comes from the knowledge of the branching fractions of the normalization modes taken
from Ref.~\cite{PDG2014} which have uncertainties of $4\%$--$21\%$, depending on the decay mode involved. The considerably
smaller uncertainty from the $\jpsi\to\mup\mun$ branching fraction is neglected.
An additional uncertainty originates from the measured value of the ratio of fragmentation fractions,
$f_{s}/f_{d}=0.259\pm 0.015$, taken from Refs.~\cite{LHCb-CONF-2013-011,LHCb-PAPER-2011-018,LHCb-PAPER-2012-037}.

As the difference in the lifetimes between the mass eigenstates of the \Bs meson, $\Delta\Gamma_s$, is significant,
the signal efficiency depends on the admixture of the \CP content of the final state~\cite{DeBruyn:2012wj}. As this is unknown for $\Bs\to\jpsi\gamma$,
two extreme scenarios are compared, where the decay is either purely \CP odd or purely \CP even. The lifetimes for the \CP
eigenstates are taken from Ref.~\cite{HFAG} to be $1.379\pm0.031$~($1.656\pm0.033$)\,ps for the \CP-even (-odd) final states. The corresponding difference in efficiency is $^{+8}_{-4}\%$ compared to the average \Bs lifetime, and is added as a systematic uncertainty.

The shape of the signal is obtained from simulation.
Potential mismodeling of this shape is assessed by comparing the signal peak position and signal width of $\Bd\to\Kstarz\gamma$ decays in data and simulation. The $\Kstarz\gamma$ invariant mass distributions
are fitted separately for long and downstream candidates using the simulation to model the signal shape and using an exponential
function to model the combinatorial background. There is no significant difference in the peak position, while the signal resolution in data is $(28\pm14)$\% and $(40\pm13)$\% wider with respect to simulation for the long and downstream categories. These factors are used to correct the signal width and are constrained in the fit.

Simulation is relied upon to model any residual kinematic differences between the signal
 and normalization channels. The ability of the simulation to
accurately emulate these differences in reconstruction is assessed by comparing simulation and data for the
$\Bd\to\Kstarz\gamma$ decay. Any differences are used to recompute the relative signal and normalization
efficiency and then assigned as systematic uncertainties. The \sPlot technique~\cite{Pivk:2004ty} is used to compare
the data and simulation for the transverse momentum of the photon
and the cosine of the pointing angle, defined as the angle between
the momentum vector of the $\Bd$ candidate and its flight direction. The effect
on the relative efficiency of reweighting the simulation to match the data is 4\% for long candidates and 2\%
for downstream candidates and these values are applied as systematic uncertainties.

\section{Results}
\label{sec:results}

The \cls method~\cite{Read:2002,Junk:1999}
is used to determine upper limits on the \BsJpsiGamma and $\Bz\to\jpsi\gamma$ branching fractions.
The test statistic used is that described in Eq.~16 of Ref.~\cite{Cowan:2007}. For a given hypothesis
of the branching fraction, $\mathcal{B}$, $q_{\mathcal{B}}$ is defined as the
ratio of likelihoods given the hypothesis value and the best-fit value,

\begin{equation}
  q_{\mathcal{B}} =
  \begin{cases}
    -2\ln\frac{\mathcal{L}(\mathrm{data}|\mathcal{B},\hat{\theta}_{\mathcal{B}})}{\mathcal{L}(\mathrm{data}|0,\hat{\theta}_{0})} & \hat{\mathcal{B}}<0 \\
    -2\ln\frac{\mathcal{L}(\mathrm{data}|\mathcal{B},\hat{\theta}_{\mathcal{B}})}{\mathcal{L}(\mathrm{data}|\hat{\mathcal{B}},\hat{\theta}_{\hat{\mathcal{B}}})} & 0\leq\hat{\mathcal{B}}\leq\mathcal{B} \\
    0 & \hat{\mathcal{B}}>\mathcal{B}
  \end{cases}
  \label{eq:teststat}
\end{equation}
where $\hat{\mathcal{B}}$ is the best-fit
branching fraction and $\hat{\theta}_{\mathcal{B}}$ are the best-fit values of the nuisance parameters
given the hypothesis value $\mathcal{B}$. Pseudoexperiments are generated in order to determine the
observed and expected exclusion confidence level of the branching fraction value. The exclusion confidence level, \cls,
is calculated as the ratio of the fraction of signal and background pseudoexperiments to the fraction
of background-only pseudoexperiments, which have a test statistic value larger than that found in data.

\begin{figure}[tb]
  \centering
  \includegraphics[width=0.49\textwidth]{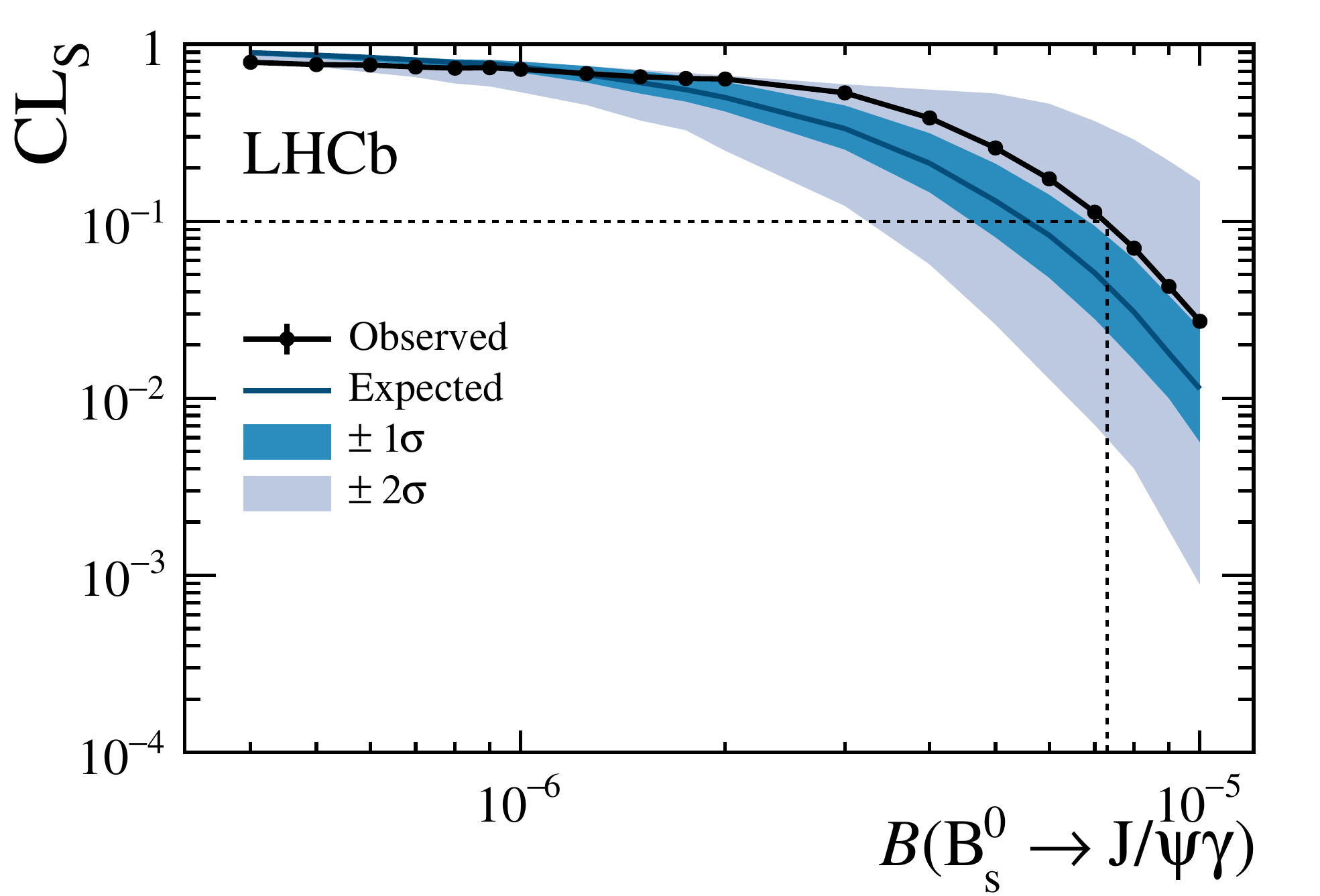}
  \includegraphics[width=0.49\textwidth]{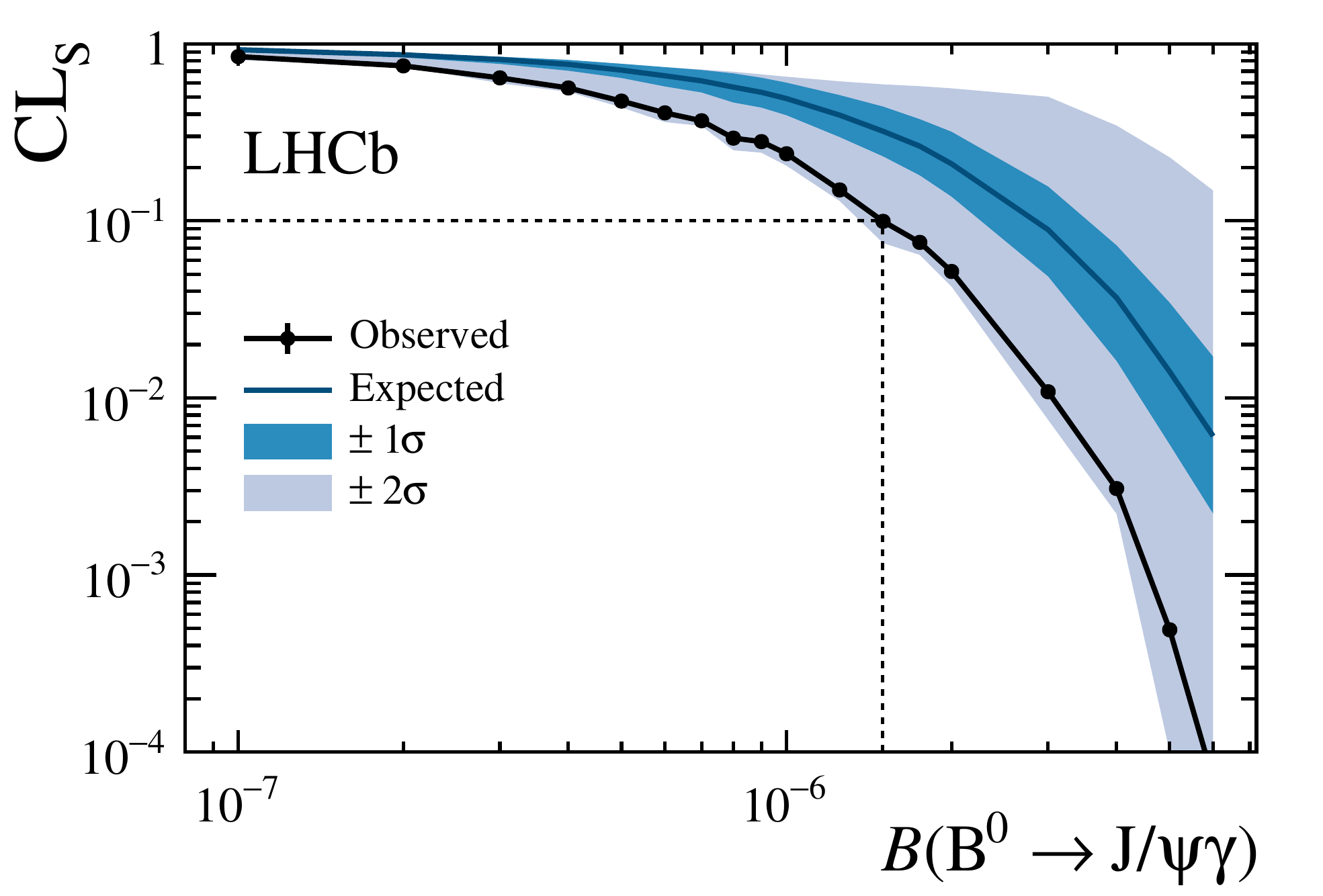}
  \caption{Observed \cls (black points), expected \cls (continuous line) and expected 1 and 2~$\sigma$ uncertainty (shaded bands) value as a function of the hypothesized branching fraction for the $\Bs\to\jpsi\gamma$ decay (left) and $\Bd\to\jpsi\gamma$ decay (right). The dashed line represents the branching fraction excluded at 90\%~C.L..}
  \label{fig:cls}
\end{figure}

The observed and expected \cls exclusions are shown as functions of the hypothesis branching fraction for \BsJpsiGamma and \BdJpsiGamma decays in Fig.~\ref{fig:cls}. The branching fraction upper limits are determined to be
\begin{displaymath}
\begin{split}
\mathcal{B}(\Bs\to\jpsi\gamma) <  7.3\;(8.7)\times10^{-6} ~\rm{at}~ 90~(95)\%~C.L., \\
\mathcal{B}(\Bz\to\jpsi\gamma) < 1.5\;(2.0)\times10^{-6} ~ \rm{at}~ 90~(95)\%~C.L.
\end{split}
\end{displaymath}

\section{Conclusion}

A search for the decays \BsJpsiGamma and $\Bz\to\jpsi\gamma$ has been performed with data collected by the LHCb experiment corresponding to an integrated luminosity of 3\invfb.
These decay modes predominantly occur via a $W$ boson exchange diagram and are sensitive to extensions of the
SM. No significant signal is observed and an upper limit on the branching fraction is set at \obslimitn at
$90\%$~C.L.~for the $\Bs\to\jpsi\gamma$ decay mode and $1.5\times10^{-6}$ at $90\%$~C.L.~for the $\Bz\to\jpsi\gamma$ decay mode.
The $\Bz\to\jpsi\gamma$ branching fraction limit is competitive with, and in agreement with, the previous measurement from $BABAR$~\cite{Aubert:2004xd}.
This is the first limit on the decay $\Bs\to\jpsi\gamma$ and is close to the sensitivity ($5\times10^{-6}$) of the calculation of the branching fraction based on perturbative QCD~\cite{Li:2006xe}.

% Do not include this in analysis note and conference reports
%\input{acknowledgements}
\section*{Acknowledgements}

\noindent We express our gratitude to our colleagues in the CERN
accelerator departments for the excellent performance of the LHC. We
thank the technical and administrative staff at the LHCb
institutes. We acknowledge support from CERN and from the national
agencies: CAPES, CNPq, FAPERJ and FINEP (Brazil); NSFC (China);
CNRS/IN2P3 (France); BMBF, DFG and MPG (Germany); INFN (Italy);
FOM and NWO (Netherlands); MNiSW and NCN (Poland); MEN/IFA (Romania);
MinES and FANO (Russia); MinECo (Spain); SNSF and SER (Switzerland);
NASU (Ukraine); STFC (United Kingdom); NSF (USA).
We acknowledge the computing resources that are provided by CERN, IN2P3 (France), KIT and DESY (Germany), INFN (Italy), SURF (The Netherlands), PIC (Spain), GridPP (United Kingdom), RRCKI (Russia), CSCS (Switzerland), IFIN-HH (Romania), CBPF (Brazil), PL-GRID (Poland) and OSC (USA). We are indebted to the communities behind the multiple open
source software packages on which we depend. We are also thankful for the
computing resources and the access to software R\&D tools provided by Yandex LLC (Russia).
Individual groups or members have received support from AvH Foundation (Germany),
EPLANET, Marie Sk\l{}odowska-Curie Actions and ERC (European Union),
Conseil G\'{e}n\'{e}ral de Haute-Savoie, Labex ENIGMASS and OCEVU,
R\'{e}gion Auvergne (France), RFBR (Russia), GVA, XuntaGal and GENCAT (Spain), The Royal Society
and Royal Commission for the Exhibition of 1851 (United Kingdom).

%\input{appendix}

% This should be taken out in the final paper
%\input{supplementary-app}

\addcontentsline{toc}{section}{References}
\setboolean{inbibliography}{true}
\bibliographystyle{LHCb}
\bibliography{main,LHCb-PAPER,LHCb-CONF,LHCb-DP,LHCb-TDR}

\newpage

% Author List ----------------------------
%  You need to get a new author list!

%\input{LHCb_HD_authorlist_2014-06-20}

\newpage
%\input{LHCb_HD_authorlist_2015-07-28.tex}
%%%%%%%%%%%%%%%%%%%%%%%%%%%%%%%%%%%%%%%%%%
\centerline{\large\bf LHCb collaboration}
\begin{flushleft}
\small
R.~Aaij$^{38}$,
C.~Abell\'{a}n~Beteta$^{40}$,
B.~Adeva$^{37}$,
M.~Adinolfi$^{46}$,
A.~Affolder$^{52}$,
Z.~Ajaltouni$^{5}$,
S.~Akar$^{6}$,
J.~Albrecht$^{9}$,
F.~Alessio$^{38}$,
M.~Alexander$^{51}$
% ...

{\footnotesize \it
% ...
$ ^{5}$Clermont Universit\'{e}, Universit\'{e} Blaise Pascal, CNRS/IN2P3, LPC, Clermont-Ferrand, France\\
$ ^{6}$CPPM, Aix-Marseille Universit\'{e}, CNRS/IN2P3, Marseille, France\\
% ...
$ ^{9}$Fakult\"{a}t Physik, Technische Universit\"{a}t Dortmund, Dortmund, Germany\\
% ...
$ ^{37}$Universidad de Santiago de Compostela, Santiago de Compostela, Spain\\
$ ^{38}$European Organization for Nuclear Research (CERN), Geneva, Switzerland\\
% ...
$ ^{40}$Physik-Institut, Universit\"{a}t Z\"{u}rich, Z\"{u}rich, Switzerland\\
% ...
$ ^{46}$H.H. Wills Physics Laboratory, University of Bristol, Bristol, United Kingdom\\
% ...
$ ^{51}$School of Physics and Astronomy, University of Glasgow, Glasgow, United Kingdom\\
$ ^{52}$Oliver Lodge Laboratory, University of Liverpool, Liverpool, United Kingdom\\
% ...
}
\end{flushleft}
%%%%%%%%%%%%%%%%%%%%%%%%%%%%%%%%%%%%%%%%%%

\end{document}